\documentclass[aps,prd,twocolumn,groupedaddress,preprintnumbers]{revtex4-2}

\usepackage{amssymb}
\usepackage{amsmath}
\usepackage{graphicx}
\usepackage{enumerate}
\usepackage{mathtools}
\usepackage{color}
\usepackage{bbold}
\usepackage{subfigure}
\usepackage{slashed}
\usepackage{xcolor}
\usepackage{bm}
\usepackage{multirow}
\usepackage{pifont}

\begin{document}

\preprint{LA-UR-25-24518}

\title{Thermodynamics of oscillating neutrinos}


\author{Lucas Johns}
\email[]{ljohns@lanl.gov}
\affiliation{Theoretical Division, Los Alamos National Laboratory, Los Alamos, NM 87545, USA}
\affiliation{Departments of Astronomy \& Physics, University of California, Berkeley, CA 94720, USA}

\begin{abstract}
The title theory is formulated. It entails a quantum-coherent variant of the Fermi--Dirac distribution and casts new light on neutrino oscillations. It might enable the incorporation of neutrino mixing into the modeling of core-collapse supernovae and neutron star mergers.
\end{abstract}

\maketitle

\section{Introduction}

Neutrino astronomy began with an unexpected deficit in the solar flux \cite{bahcall1976}. The puzzle's resolution finally arrived three decades later, with experimental confirmation of flavor oscillations \cite{davis2003, koshiba2003, kajita2016, mcdonald2016}. Advances in the theory of neutrino propagation were integral to this triumph \cite{gribov1969, nussinov1976, wolfenstein1978, mikheyev1985, bethe1986, haxton1986, parke1986, kuo1989}.

Neutrino transport theory is today in need of another wave of progress. The urgency comes especially from core-collapse supernovae and neutron star mergers \cite{johns2025neutrino}. Even though these sites are two of the marquee targets of multimessenger astronomy, and two of the most carefully modeled systems in computational astrophysics, neutrino oscillations are yet to be reliably incorporated into the relevant simulations and predictions \cite{janka2016, shibata2019, mezzacappa2020, muller2020, burrows2021, baiotti2017, radice2020, metzger2020}. In recent years, evidence has piled up that the current situation is unacceptable. The best estimates of flavor mixing's effects---on explosion dynamics, nucleosynthesis, kilonova light curves, and emitted neutrino signals---point to them being substantial \cite{mirizzi2016, wu2017b, george2020, xiong2020, li2021, nagakura2021c, just2022, fernandez2022, johns2023, xiong2022b, xiong2023, nagakura2023, ehring2023b, ehring2023}.

Efforts to solve the oscillation problem in compact stellar environments have been based on kinetic theory \cite{notzold1988, pantaleone1992, sigl1993, raffelt1993, raffelt1993b, loreti1994, yamada2000, friedland2003, strack2005, cardall2008, duan2010, volpe2013, vlasenko2014, kartavtsev2015, stirner2018, nagakura2022b, johns2023b}. The challenge is that the equations are nonlinear, multiscale, and far beyond the reach of direct numerical simulation. This sounds dire, but an analogy gives us hope. Climate simulations would likewise be out of reach if they were based on the Boltzmann equation for air and water molecules. And yet, because of hydrodynamics, sophisticated modeling is possible.

\begin{figure}
\centering
\includegraphics[width=.43\textwidth]{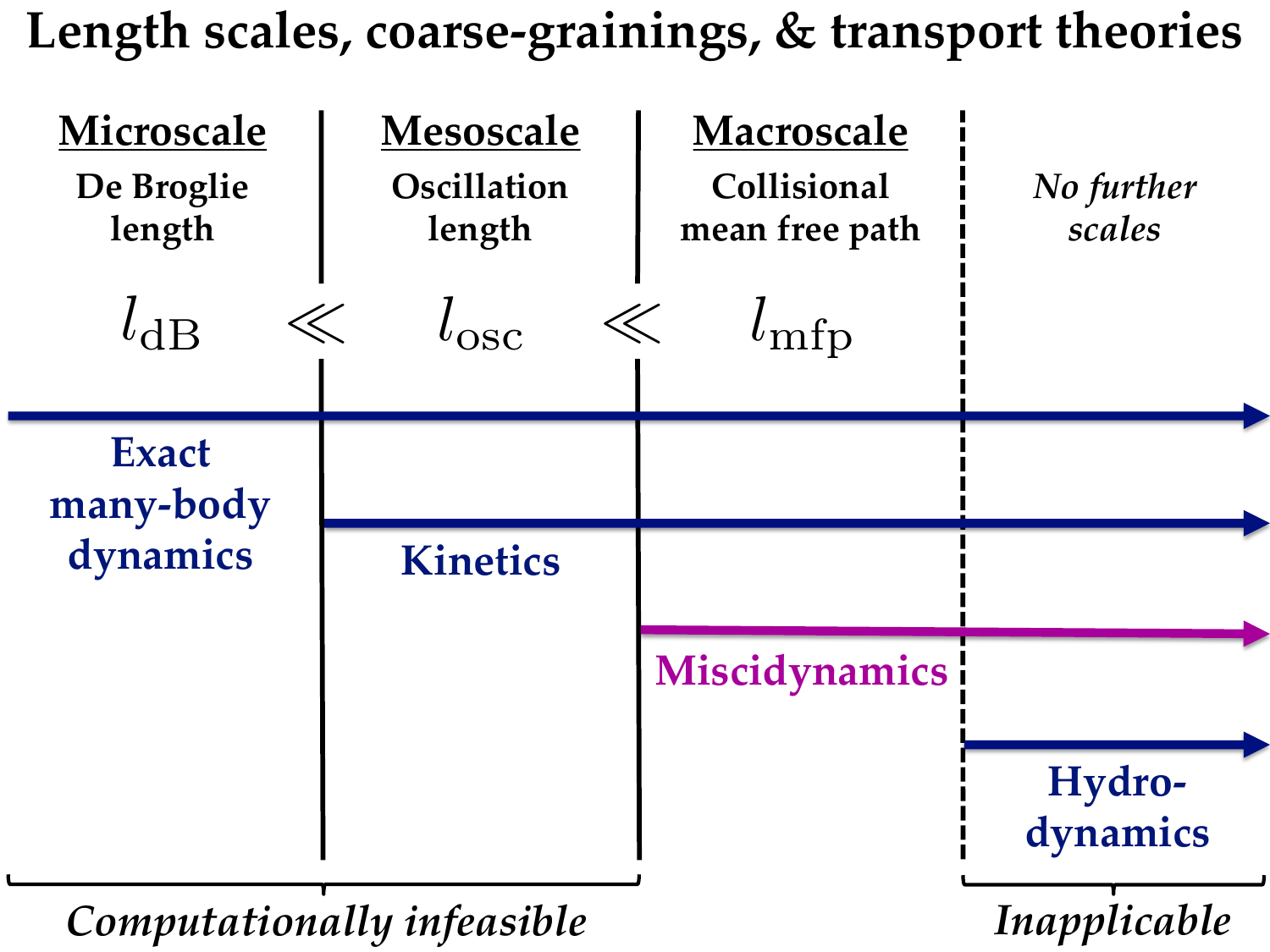}
\caption{In core-collapse supernovae and neutron star mergers, neutrino oscillations unfold at the mesoscale ($10^{-3}~\textrm{m} \lesssim l_\textrm{osc} \lesssim 1~\textrm{m}$), well-separated from the quantum microscale ($l_\textrm{dB} \lesssim 10^{-14}~\textrm{m}$) and the collisional macroscale ($l_\textrm{mfp} \gtrsim 10^3~\textrm{m}$). Kinetics is unworkable because it attempts to resolve the mesoscopic details. Hydrodynamics is as well because optical depths are low. Miscidynamics is a new macroscopic transport theory based on local mixing equilibrium.}
\label{fig:cging}
\end{figure}

For neutrino atmospheres, hydrodynamics is too coarse. Oscillations occur where neutrinos are at most weakly collisional. The appropriate theory, which we aim to develop, is intermediate between kinetics and hydrodynamics (Fig.~\ref{fig:cging}). We call it \textit{miscidynamics}, borrowing the Latin root shared with words like \textit{mixing} and \textit{miscible}, because it describes transport under the condition that mixing is in local equilibrium. Kinetics coarse-grains over the microscopic dynamics at the de Broglie scale but stops short of oscillations. This level of detail is unnecessary and intractable. In miscidynamics, the granularity is set by the same macroscopic scales that determine the numerical resolution in simulations.

En route to miscidynamics, we first need to formulate the concept of mixing equilibrium. The subject of particle mixing dates back nearly 70 years \cite{gellmann1955, pontecorvo1957, feinberg1961}. It has far-reaching significance in the Standard Model and beyond. Nevertheless, a thermodynamic theory of the phenomenon has not yet been proposed. Only now do we face an open problem that seems to require one.

The new thermodynamics is interesting in its own right, apart from its possible utility for transport. It describes the equilibrium phases of collisionless neutrino matter, \textit{i.e.}, of relativistic fermions that exist indefinitely in mass superpositions. Thermalization occurs without any dissipation or decoherence, facilitated instead by small-scale fluctuations of the neutrino system itself. Below, we derive the equilibrium theory from kinetics and discuss some of its basic aspects. We then broaden the scope of the theory to encompass transport.

Overall, the viewpoint here is that neutrino mass raises fundamental questions not only for particle physics (which it does by demanding content beyond the Standard Model) but for statistical physics as well.

\section{Kinetics}

Our starting point is the kinetic equation
\begin{align}
i \left( \partial_t + \boldsymbol{\hat{p}} \cdot \partial_{\boldsymbol{x}} \right) \rho_\nu \left( t, \boldsymbol{x}, \boldsymbol{p} \right) = &\left[ H_\nu (t, \boldsymbol{x}, \boldsymbol{p} ), \rho_\nu \left( t, \boldsymbol{x}, \boldsymbol{p} \right) \right] \notag \\
&~~~~~~~~ + iC_\nu (t, \boldsymbol{x}, \boldsymbol{p}). \label{eq:qke}
\end{align}
The density matrix $\rho_\nu$ is an $N_f \times N_f$ matrix, where $N_f$ is the number of flavors. We take $N_f = 2$ throughout this work. Quantities for antineutrinos will be denoted by the same symbols but with subscript $\bar{\nu}$. The diagonal components of $\rho_\nu$ are distribution functions in the chosen basis. Off-diagonal components measure quantum coherence. In the thermodynamics of \textit{massless} neutrinos, the equilibrium distribution $\rho^\textrm{eq}$ is diagonal in the flavor basis with entries $f_\textrm{FD} (p, T, \mu_{\nu_\alpha})$, where $p \equiv |\boldsymbol{p}|$. That is, each flavor $\alpha$ ($ = e, \mu, \tau$) has a Fermi--Dirac distribution at temperature $T$ and chemical potential $\mu_{\nu_\alpha}$.

The evolution is dictated by the Hamiltonian $H_\nu$ and collisional term $iC_\nu$. The latter represents Boltzmann integrals for all relevant processes. We will not be concerned with the particulars. The Hamiltonian, on the other hand, is crucial for our purposes because it is responsible for mixing equilibration. We split it into $H_\nu = H_\nu^{(1)} + H_\nu^{(2)}$. The one-body part is
\begin{equation}
H_\nu^{(1)} \left( t, \boldsymbol{x}, \boldsymbol{p} \right) = p + \frac{M^2}{2 p} + \sqrt{2} G_F \left( 1 - \boldsymbol{v}_m ( t, \boldsymbol{x} ) \cdot \boldsymbol{\hat{p}} \right) L (t, \boldsymbol{x}), \label{eq:Hnu1}
\end{equation}
where $M^2$ is diagonal in the mass basis with components $m_i^2$, $L$ is diagonal in the flavor basis with components $n_{\alpha^-} - n_{\alpha^+}$, $\boldsymbol{v}_m$ is the velocity of background matter, and $G_F$ is the Fermi constant. The terms on the right-hand side are the neutrino momentum, the vacuum-oscillation Hamiltonian, and the potential generated by neutrino refraction on matter. $H_{\bar{\nu}}^{(1)}$ differs only in that the $p$ and $M^2$ terms have minus signs in front of them. The two-body part of the Hamiltonian is
\begin{equation}
H_{\nu}^{(2)} \left( t, \boldsymbol{x}, \boldsymbol{p} \right) = \sqrt{2} G_F \left( D_0 \left( t, \boldsymbol{x} \right) - \boldsymbol{\hat{p}} \cdot \boldsymbol{D}_1 \left( t, \boldsymbol{x} \right) \right), \label{eq:Hnu2}
\end{equation}
where we use the notation
\begin{equation}
D_l \left( t, \boldsymbol{x} \right) = \int \frac{d^3 \boldsymbol{q}}{(2\pi)^3} \left( \boldsymbol{\hat{q}} \right)^l \left( \rho_\nu \left( t, \boldsymbol{x}, \boldsymbol{q} \right) - \rho_{\bar{\nu}} \left( t, \boldsymbol{x}, \boldsymbol{q} \right) \right). \label{eq:Dmatrices}
\end{equation}
It contains the nonlinearity due to neutrino--neutrino refraction and is identical for antineutrinos.

The total energy is $U = \textrm{Tr} \int d^3\boldsymbol{x}~u(\boldsymbol{x})$ with
\begin{equation}
u = \int \frac{d^3 \boldsymbol{p}}{(2\pi)^3} \left( H_{\nu}^{(1)} \rho_\nu - H_{\bar{\nu}}^{(1)} \rho_{\bar{\nu}}  \right) + \frac{\sqrt{2} G_F}{2} \left( D_0^2 - \boldsymbol{D}_1^2 \right). \label{eq:uenergy}
\end{equation}
It is strictly conserved when the system is spatially homogeneous with $iC_{\nu} = 0$ \cite{fiorillo2024}. In addition, $\textrm{Tr}[\rho^{n}_\nu (t, \boldsymbol{x}_0 + \boldsymbol{\hat{p}} t, \boldsymbol{p})]$ is time-invariant for given $\boldsymbol{x}_0$ and $\boldsymbol{p}$ and for $1 \leq n \leq N_f$. These invariants enforce conservation of neutrino number ($n = 1$) and flavor polarization ($n = 2$) along each phase-space trajectory. Their invariance constitutes a sort of quasiclassical Liouville's theorem that forbids true, fine-grained equilibration. Thermalization occurs only at a coarse-grained level, through the formation of small-scale structure in phase space, as in collisionless classical systems \cite{johns2020b}. Indeed, the agenda we pursue in this paper has parallels with the one initiated by Lynden-Bell for gravitational systems \cite{lynden1967}.

\section{Entropy and equilibrium}

To prepare for coarse-graining, we put our neutrino system in a periodic box of volume $V = l_\textrm{box}^3$ with side length satisfying $l_\textrm{osc} \ll l_\textrm{box} \lesssim l_\textrm{astro}$, where $l_\textrm{astro}$ is the length scale on which the medium varies. The purpose of $l_\textrm{osc} \ll l_\textrm{box}$ is to ensure that neutrinos reach mixing equilibrium before the boundaries of the box become relevant. Very roughly, $l_\textrm{eq} \sim l_\textrm{osc}$. The purpose of $l_\textrm{box} \lesssim l_\textrm{astro}$ is to ensure that all parameters apart from the neutrino distributions are approximately homogeneous. Later, when we turn our attention to transport, we will be stitching these boxes together to construct a global description of the neutrino radiation. Ideally, it will be possible to take $l_\textrm{box} = l_\textrm{sim}$, the spatial resolution that would be used in a simulation \textit{without} neutrino oscillations.

We now perform a spatial average over the box, making any smaller-scale information inaccessible. This is not just a useful thing to do. It is realistic. The only way the information could become known is through interactions with the astrophysical fluid. But the fluid is only sensitive to features on the weak-interaction scale $l_\textrm{mfp}$, and the neutrino--fluid coupling is not strong in the regions we are considering ($l_\textrm{mfp} \gtrsim l_\textrm{box}$). The coarse-graining is imposed by nature itself.

The logic above also applies to temporal resolution. We take $t_\textrm{osc} \ll t_\textrm{box} \lesssim t_\textrm{astro}$ because it is both practical and necessary. Our next step is to assume that the evolution within the box is ergodic. This is a plausible hypothesis given that collective oscillations are known to exhibit instabilities, chaos, quasi-steady states, and so on. Granting ergodicity, we are able to make our first contact with thermodynamics. Instead of attempting to describe the time- and space-averaged evolution, we calculate coarse-grained expectation values with respect to the ensemble of all fine-grained states that the system is able to visit.

To carry out this program, we need an entropy. It should be a functional of $\overline{\rho_{\nu,\boldsymbol{p}}}$, the $\boldsymbol{x}$-average of $\rho_{\nu,\boldsymbol{x},\boldsymbol{p}}$. The change to subscripts for $\boldsymbol{x}$ and $\boldsymbol{p}$ is meant to emphasize the fact that we are now regarding $\rho_\nu$ not as a time-dependent solution but rather as a microstate in a statistical ensemble. Because neutrinos are fermions, the appropriate entropy is
\begin{equation}
S = V \int \frac{d^3 \boldsymbol{p}}{(2\pi)^3} \left( s_{\nu,\boldsymbol{p}} + s_{\bar{\nu},\boldsymbol{p}} \right) \label{eq:entropy1}
\end{equation}
with
\begin{equation}
s_{\nu,\boldsymbol{p}} = -\textrm{Tr} \left[ \overline{\rho_{\nu,\boldsymbol{p}}} \log \overline{\rho_{\nu,\boldsymbol{p}}} + \left( 1 - \overline{\rho_{\nu,\boldsymbol{p}}} \right) \log \left( 1 - \overline{\rho_{\nu,\boldsymbol{p}}} \right) \right]. \label{eq:entropy2}
\end{equation}
The matrix log function is defined by diagonalizing $\overline{\rho_{\nu,\boldsymbol{p}}}$ and applying the scalar log function to the diagonal elements. Using a similar formula for the fine-grained system would result in $S = \textrm{constant}$ because the entropy is determined by the fine-grained invariants $\textrm{Tr}\left[ \rho_\nu^n \right]$. Coarse-graining preserves neither the $n=2$ trace nor the energy $U$ and thus enables effective decoherence and dissipation.

Although $U[\overline{\rho_{\nu,\boldsymbol{p}}}]$ need not be strictly equal to $\mathcal{U} \equiv \overline{U[\rho_{\nu,\boldsymbol{x},\boldsymbol{p}}]}$, in equilibrium we expect them to differ only as a result of thermal fluctuations. We therefore maximize $S$ subject to the constraint that $U[\overline{\rho_{\nu,\boldsymbol{p}}}] \cong \mathcal{U}$. To do this, we introduce a Lagrange multiplier $\beta$. Conceptually, the coarse-grained variables are in thermal contact with a heat bath of temperature $T \equiv \beta^{-1}$ that consists of the system's own fine-grained fluctuations.

The neutrino number $N_{\nu,\boldsymbol{p}}$ is exactly conserved at the coarse-grained level because $N_{\nu,\boldsymbol{p}} = V \overline{\textrm{Tr}\left[ \rho_{\nu,\boldsymbol{x},\boldsymbol{p}} \right]} = V \textrm{Tr}\left[ \overline{\rho_{\nu, \boldsymbol{p}}} \right]$. Depending on the application, there may be other conservation laws. To indicate how the analysis generalizes, we take the set of conserved quantities to be $\lbrace U, N_{\nu,\boldsymbol{p}}, N_{\bar{\nu},\boldsymbol{p}}, G \rbrace$ with Lagrange multipliers $\lbrace \beta, \eta_{\nu,\boldsymbol{p}}, \eta_{\bar{\nu},\boldsymbol{p}}, \lambda \rbrace$. Here $G$ is some unspecified scalar quantity. All Lagrange multipliers are scalars as well. Continuing to use calligraphic letters for expectation values, the entropy with constraints is
\begin{align}
S' \equiv S &+ \beta \left( U - \mathcal{U} \right) - \int \frac{d^3 \boldsymbol{p}}{(2\pi)^3} \big( \eta_{\nu,\boldsymbol{p}} \left( N_{\nu,\boldsymbol{p}} - \mathcal{N}_{\nu,\boldsymbol{p}} \right) \notag \\
&+ \eta_{\bar{\nu},\boldsymbol{p}} \left( N_{\bar{\nu},\boldsymbol{p}} - \mathcal{N}_{\bar{\nu},\boldsymbol{p}} \right) \big) + \lambda \left( G - \mathcal{G} \right).
\end{align}
Then, from the extremization condition $\delta S' / \delta \overline{\rho_{\nu,\boldsymbol{p}}}|_{\rho^\textrm{eq}_{\nu,\boldsymbol{p}}} = 0$, we obtain the mixing-equilibrium distribution:
\begin{equation}
\rho^\textrm{eq}_{\nu,\boldsymbol{p}} = \frac{1}{\exp \left[ \beta \left( H^\textrm{eq}_{\nu,\boldsymbol{p}} - \mu_{\nu,\boldsymbol{p}}  \right) + \lambda \left( \delta G / \delta \overline{\rho_{\nu,\boldsymbol{p}}} \right) \right] + 1}. \label{eq:rhoeq}
\end{equation}
We have defined $\mu_{\nu,\boldsymbol{p}} \equiv \eta_{\nu,\boldsymbol{p}} / \beta$ and $H^\textrm{eq}_{\nu,\boldsymbol{p}} \equiv H_{\nu,\boldsymbol{p}} [ \rho^\textrm{eq} ]$. The right-hand side is a matrix function defined through its power series. Scalar terms inside the matrix exponential, such as $\beta\mu_{\nu,\boldsymbol{p}}$, as well as the scalar $1$ in the denominator, are implicitly multiplied by the $N_f \times N_f$ identity matrix. Supposing that the invariant $G$ is linear in $\rho$, the $\lambda$ term is proportional to the identity matrix and the argument of the exponential is Hermitian. The right-hand side can then be calculated by diagonalization, applying the function to each of the eigenvalues of the argument. Explicit evaluation of $\rho^\textrm{eq}$ generally requires the solution of self-consistency conditions, a point we will return to.

\section{Work and heat}

With the concepts of entropy and equilibrium defined, other devices in the thermodynamic apparatus become available. We could, for example, make use of the coarse-grained free energy $F = U - TS$. Most importantly, we need to consider how neutrinos move between and into equilibria. The indispensable concepts in this respect are work and heat. 

To define them, we use the Pauli matrices $\sigma_a$ to decompose $\rho_\nu$ and $H_\nu$ into scalar and vector parts:
\begin{equation}
\rho_\nu = \frac{P_{\nu,0} + \vec{P}_\nu \cdot \vec{\sigma}}{2}, ~~~ H_\nu = \frac{H_{\nu,0} + \vec{H}_\nu \cdot \vec{\sigma}}{2}.
\end{equation}
The first law of thermodynamics is then (dropping subscripts and integrals for readability)
\begin{align}
\Delta U = &\overbrace{\frac{1}{2} H_0 \Delta P_0 + \frac{1}{2} \vec{H} \cdot \Delta| \vec{P}| \hat{P}}^{~ \equiv Q^\textrm{env}} + \overbrace{\frac{1}{2} | \vec{H} | | \vec{P} | \Delta \left( \hat{H} \cdot \hat{P} \right)}^{~ \equiv Q^\textrm{kin}} \notag \\
&~~~~~~~~~~ +\underbrace{\frac{1}{2} \Delta H_0 P_0 + \frac{1}{2} \Delta | \vec{H} | | \vec{P} | \hat{H} \cdot \hat{P}}_{\equiv  W}. \label{eq:firstlaw}
\end{align}
The upper line is the heat $Q$ gained by the system during some process and the lower line is the work $W$ done on it. $Q^\textrm{env}$ can be collisionally transferred from the medium or internally generated by neutrino--neutrino collisions; the environment in the latter case is the fluctuating bath of neutrino many-body correlations \cite{johns2023b, kost2024once}. The literature on equilibration in isolated quantum systems has emphasized heating essentially of this second type, with thermal behavior emerging as entanglement spreads throughout the system \cite{gogolin2016, dalessio2016, ueda2020}. For neutrino mixing, the predominant equilibration mechanism is \textit{kinematic} heating $Q^\textrm{kin}$ caused by dealignment of $\hat{P}$ with $\pm \hat{H}$.

We may as well state the other laws of thermodynamics at this stage. The second law is $S[\rho^\textrm{eq}] \geq S[\rho^\textrm{in}]$ for any initial state $\rho^\textrm{in}$. We take it as axiomatic, but presumably one could prove a neutrino $H$-theorem by writing out the infinite BBGKY hierarchy of $\overline{\rho_1 \dotsb \rho_m}$ correlators and truncating via molecular chaos. This calculation would be formally similar to (but physically very different from) the derivation of kinetics from the BBGKY hierarchy of quantum expectation values \cite{volpe2013, froustey2020}. The result would be a new type of kinetic equation in which collisions occur between coarse-grained flavor fields rather than individual particles. It could perhaps be used to formulate viscous miscidynamics.

The third law follows from using Eq.~\eqref{eq:rhoeq} to identify the unique ground state:
\begin{equation}
\left( \rho^\textrm{eq}_{\nu,\boldsymbol{p}} \right)_{IJ} \xrightarrow{T \rightarrow 0}  \begin{cases}
\delta_{IJ} & \left( H_{\nu,\boldsymbol{p}}^\textrm{eq} \right)_{IJ} \leq \mu_{\nu,\boldsymbol{p}} \\ 
0 & \left( H_{\nu,\boldsymbol{p}}^\textrm{eq} \right)_{IJ} > \mu_{\nu,\boldsymbol{p}}, \label{eq:degenerate}
\end{cases}
\end{equation}
written in the basis that diagonalizes $H_{\nu,\boldsymbol{p}}^\textrm{eq}$. A fully occupied or vacant level has no entropy, and therefore $S = 0$ for the system as a whole. Interestingly, for fixed $\boldsymbol{\hat{p}}$ and $\boldsymbol{p}$-independent $\mu_{\nu,\boldsymbol{p}}$, there is a very narrow $p$ band in which the levels are a hybrid of fully occupied and fully vacant. This is because the Fermi surface in $\boldsymbol{p}$-space is mass-/flavor-dependent.

Later we will use our definitions of work and heat to interpret oscillation phenomena as thermodynamic processes.

\section{Miscidynamics\label{sec:misc}}

We are now in a position to outline the application of the equilibrium theory to astrophysical neutrino transport. Here we present miscidynamics only in the adiabatic limit, leaving the derivation of nonadiabatic corrections to future work. 

We will state the result first, then derive it. Applying the condition of local mixing equilibrium to Eq.~\eqref{eq:qke}, and enforcing kinematic adiabaticity ($Q^\textrm{kin} = 0$) at every step, we obtain the miscidynamic equation in the adiabatic limit:
\begin{equation}
i \left( \partial_t + \boldsymbol{\hat{p}} \cdot \partial_{\boldsymbol{x}} \right) \rho_\nu^\textrm{eq} (t, \boldsymbol{x}, \boldsymbol{p}) = iC_{\nu,\textrm{non}}^\textrm{eq} (t, \boldsymbol{x}, \boldsymbol{p}). \label{eq:misc}
\end{equation}
This form applies in the $t$-, $\boldsymbol{x}$-, and $\boldsymbol{p}$-dependent frame in which $H_\nu^{\textrm{eq}}$ is everywhere and always diagonal. The collision integrals are those in Eq.~\eqref{eq:qke}, but evaluated using $\rho_\nu^\textrm{eq}$ and discarding the part that causes a change in $\hat{P}_{\nu}$. That is, only the instantaneously nonunitary part of $iC_\nu$ is retained because the unitary part is absorbed into the frame transformation. The collision term generally includes flavor-off-diagonal elements \cite{blaschke2016, richers2019}. The commutator with the Hamiltonian vanishes. We can regard $(t, \boldsymbol{x}, \boldsymbol{p})$ as either the coarse-grained variables of a specific kinetic solution or the fine-grained variables of an ensemble average.

Now let us derive the adiabatic miscidynamic equation [Eq.~\eqref{eq:misc}] by coarse-graining the neutrino quantum kinetic equation [Eq.~\eqref{eq:qke}]. First we define the coarse-graining operator 
\begin{equation}
\left\langle \cdot \right\rangle (T, \boldsymbol{X}, \boldsymbol{p}) \equiv \frac{1}{V \Delta t} \int_{\mathcal{R}_{\boldsymbol{x}}} d^3\boldsymbol{x}' \int_t^{t+\Delta t} dt' \left( \cdot \right) \left( t', \boldsymbol{x}', \boldsymbol{p} \right), \label{eq:cgop}
\end{equation}
which averages over a region $\mathcal{R}_{\boldsymbol{x}}$ with volume $V$ centered at location $\boldsymbol{x}$ and over a time step $\Delta t$. We are not coarse-graining momentum $\boldsymbol{p}$, though nothing prevents us from doing so. The use of $T = t$ and $\boldsymbol{X} = \boldsymbol{x}$ on the left-hand side emphasizes that these are the coordinates of coarse-grained variables. (Elsewhere in the paper we use $T$ to denote temperature. In this section only we use it as a time coordinate.) Thinking of the coordinates themselves as being coarse-grained is justified by the idempotent property
\begin{equation}
\left\langle \left\langle \cdot \right\rangle \right\rangle \cong \left\langle \cdot \right\rangle \label{eq:idem}
\end{equation}
of the coarse-graining operator. We can therefore regard $\langle \cdot \rangle$ as acting on the fine-grained $t$ and $\boldsymbol{x}$ but seeing the coarse-grained $T$ and $\boldsymbol{X}$ as constants. Eq.~\eqref{eq:idem} holds as long as we take $\Delta t$ and $V$ to be small relative to the scales of global variations (but still large enough that small-scale features are smoothed out, otherwise we gain nothing from the operator). In the application to $\rho_\nu$, we are taking $\Delta t \sim t_\textrm{box}$ and $V \sim l_\textrm{box}^3$ and using the scale separations $t_\textrm{osc} \ll t_\textrm{astro}$ and $l_\textrm{osc} \ll l_\textrm{astro}$.

We now introduce a frame transformation $U_{\nu} (T, \boldsymbol{X}, \boldsymbol{p})$ that diagonalizes the coarse-grained Hamiltonian:
\begin{equation}
    H^d_{\nu} = U^\dagger_{\nu} \langle H_{\nu} \rangle U_{\nu},
\end{equation}
with $H^d_{\nu} (T, \boldsymbol{X}, \boldsymbol{p})$ diagonal. After applying the frame transformation, Eq.~\eqref{eq:qke} becomes
\begin{equation}
    i ( \partial_t + \boldsymbol{\hat{p}}\cdot\partial_{\boldsymbol{x}}) \rho^d_{\nu} = [ H^d_{\nu} + \mathcal{C}^d_{\nu}, \rho^d_{\nu}] + iC^d_{\nu}, \label{eq:qked}
\end{equation}
where $\rho^d_{\nu}$ and $iC^d_{\nu}$ are the frame-transformed $\rho_{\nu}$ and $iC_{\nu}$, respectively. We assume that the nonadiabatic part of the effective Hamiltonian vanishes:
\begin{equation}
    \mathcal{C}^d_{\nu} \equiv -i U_{\nu}^\dagger ( \partial_t + \boldsymbol{\hat{p}}\cdot\partial_{\bm{x}}) U_{\nu}  \cong 0. \label{eq:nonadiab0}
\end{equation}
We are here making the assumption that coarse-grained quantities, and therefore $U_{\nu}$, vary infinitely slowly. The approach we are taking in this derivation is to assume that there exists a slowly varying $U_{\nu}$ capable of absorbing the coarse-grained unitary evolution as $\langle \rho_{\nu} \rangle$ tracks local mixing equilibrium. We can then use self-consistency to infer what $U_{\nu}$ must be.

Our next step is to write
\begin{equation}
\rho^d_\nu (t, \boldsymbol{x}, \boldsymbol{p}) = \left\langle \rho^d_\nu \right\rangle (T, \boldsymbol{X}, \boldsymbol{p})  + \delta \rho^d_\nu (t, \boldsymbol{x}, \boldsymbol{p}), \label{eq:miscexpand}
\end{equation}
which defines the deviation $\delta \rho^d_\nu$ from the average. We have
\begin{equation}
\left\langle \delta \rho^d_\nu \right\rangle \cong 0, \label{eq:miscavgdelta}
\end{equation}
consistent with Eq.~\eqref{eq:idem}. Now we use Eqs.~\eqref{eq:nonadiab0} and \eqref{eq:miscexpand} in Eq.~\eqref{eq:qked} and act on the resulting equation with $\left\langle \cdot \right\rangle$. Terms with no factors of $\delta \rho^d_\nu$ are unchanged according to Eq.~\eqref{eq:idem}. Terms that are linear in $\delta \rho^d_\nu$ vanish by Eq.~\eqref{eq:miscavgdelta}. The remaining terms contain correlators like $\left\langle \delta \rho^d_\nu \delta \rho^d_\nu \right\rangle$, $\left\langle \delta \rho^d_\nu \delta \rho^d_{\bar{\nu}} \right\rangle$, and so on. All $\delta \rho^d_\nu$ and $\delta \rho^d_{\bar{\nu}}$ factors are at the same $t$ and $\boldsymbol{x}$ but can have different $\boldsymbol{p}$.

By the ergodic hypothesis,
\begin{equation}
\left\langle \rho^d_\nu \right\rangle (T, \boldsymbol{X}, \boldsymbol{p}) \cong \overline{\rho_{\nu,\boldsymbol{p}}} (T, \boldsymbol{X}). \label{eq:miscavgerg}
\end{equation}
The left-hand expression is the time- and position-averaged dynamical solution at $(T, \boldsymbol{X})$, as defined in Eq.~\eqref{eq:cgop}. The right-hand expression is the ensemble average of possible microstates of the $iC_\nu = 0 $ system with parameters (\textit{e.g.}, electron density) set to the values at $(T, \boldsymbol{X})$. The implication of Eq.~\eqref{eq:miscavgerg} is that $\left\langle \delta \rho^d_\nu \delta \rho^d_\nu \right\rangle$ and other such correlators describe the statistics of fluctuations around the ensemble average.

To impose local mixing equilibrium, we identify
\begin{equation}
\overline{\rho_{\nu,\boldsymbol{p}}} (T, \boldsymbol{X}) = \rho_\nu^\textrm{eq} (T, \boldsymbol{X}, \boldsymbol{p}),
\end{equation}
where $\rho_\nu^\textrm{eq}$ refers to the equilibrium distribution satisfying
\begin{equation}
[H^{\textrm{eq}}_{\nu}, \rho^{\textrm{eq}}_{\nu}] = 0.
\end{equation}
The coordinates specify when and where the parameters are drawn that enter into the self-consistent determination of $H^{\textrm{eq}}_{\nu}$ and $\rho^{\textrm{eq}}_{\nu}$ [Eq.~\eqref{eq:selfcons}]. Self-consistency of local equilibrium also requires that we make the replacement
\begin{equation}
    iC^d_\nu \longrightarrow iC^{\textrm{eq}}_{\nu,\textrm{non}},
\end{equation}
where $iC^{\textrm{eq}}_{\nu,\textrm{non}}$ is obtained from $iC^d_\nu$ by setting the off-diagonal terms to zero. The justification for this step is that the instantaneous unitary part $iC^{\textrm{eq}}_{\nu,\textrm{uni}}$ is already accounted for by $U_{\nu}$, leaving the non-unitary remainder $iC^{\textrm{eq}}_{\nu,\textrm{non}}$. In local equilibrium, $\hat{P}^d_{\nu} = \pm\hat{z}$, so that the diagonal and off-diagonal parts of the collision term correspond to non-unitary and unitary changes, respectively.

Let us assume that fluctuations are either negligible or random, such that all of these correlation terms go to zero. After switching to ensemble averages, dropping fluctuation correlations, and reverting to symbols $\boldsymbol{t}$ and $\boldsymbol{x}$, we arrive at Eq.~\eqref{eq:misc}. The solution can be found following the procedure described below. The transformation $U_{\nu}$, which is implicit in Eq.~\eqref{eq:misc}, can then be inferred from the solution.

Under Eq.~\eqref{eq:misc}, all oscillation phenomena disappear into the ``eq'' superscripts with the lone exception of collisional instabilities \cite{johns2023, johns2022, johns2022b, padillagay2022b, lin2023, xiong2023, xiong2022b, liu2023, johns2023d, kato2024collisional}. This peculiar quality of the latter was foreshadowed in the analysis of Ref.~\cite{johns2023}: the oscillation terms ensure adiabaticity (``synchronized motion'') but otherwise drop out of the equations. Notably, in adopting the assumption of kinematic adiabaticity, we are ignoring any oscillation effects arising from subgrid correlations or any switching of the sign of $\hat{P}^{\textrm{eq}}_\nu \cdot \hat{H}^{\textrm{eq}}_\nu$ \cite{johns2024}. The changing medium causes $W \neq 0$, and collisions cause $Q^\textrm{env} \neq 0$, but local mixing equilibrium is tracked without any heating by oscillations.

Supposing that we have $( \rho_\nu )_i = ( \rho_\nu^\textrm{eq} )_i$ at time $t_i$, the next step in a simulation is then taken in two parts:

(1) Evolve forward to $t_{i+1}$ in the usual manner but using Eq.~\eqref{eq:misc} in place of the neutrino Boltzmann equation.

(2) Equilibrate $( \rho_\nu )_{i+1}$ by self-consistently imposing $Q^\textrm{kin} = 0$. This involves rotating each polarization vector $( \vec{P}_\nu )_{i+1}$ such that it remains (anti)aligned with $( \vec{H}_\nu )_{i+1}$.

In step (2), self-consistency conditions need to be solved because the polarization vectors are coupled to one another. Fortunately, the parts of $( \vec{H}_\nu )_{i+1}$ that are not prescribed by the medium are limited to $( \vec{D}_0 )_{i+1}$ and $( \vec{\boldsymbol{D}}_1 )_{i+1}$, the vectors associated with $D_0$ and $\boldsymbol{D}_1$ [Eq.~\eqref{eq:Dmatrices}]. The self-consistency conditions are therefore
\begin{equation}
\vec{D}_l = \int \frac{d^3 \boldsymbol{p}}{(2\pi)^3} \left( \boldsymbol{\hat{p}} \right)^l \left( s_\nu | \vec{P}_\nu | \hat{H}_\nu - s_{\bar{\nu}} | \vec{P}_{\bar{\nu}} | \hat{H}_{\bar{\nu}} \right), \label{eq:selfcons}
\end{equation}
where $s_\nu \equiv \hat{H}_\nu \cdot \hat{P}_\nu = \pm 1$. Functions in the integrand depend on $\boldsymbol{x}$ and $\boldsymbol{p}$, and all quantities are evaluated at $t_{i+1}$. The alignment factors $s_\nu$ remain constant from $t_i$ to $t_{i+1}$. The magnitudes $| \vec{P}_\nu |$ are determined from the output of step (1). $\vec{D}_0$ and $\vec{\boldsymbol{D}}_1$ appear implicitly on the right-hand side through $\hat{H}_\nu$.

From the simulator's perspective, implementing adiabatic miscidynamics comes down to promoting distribution functions to density matrices, calculating matrix-structured collisional terms, and solving Eq.~\eqref{eq:selfcons} at each location.

\section{Examples}

In this section we present several applications of miscidynamics and neutrino quantum thermodynamics. We group the examples under the categories of adiabaticity (\textit{i.e.}, adiabatic tracking of mixing equilibrium as described by adiabatic miscidynamics) and equilibration (\textit{i.e.}, situations in which neutrinos move toward mixing equilibrium from initially out-of-equilibrium states).

\subsection{Adiabaticity}

As a neutrino moves from one region to another, it experiences the changing parameters of the medium. Some of the most notable flavor-mixing phenomena result from this kind of parametric variation: Mikheyev--Smirnov--Wolfenstein (MSW) resonances \cite{wolfenstein1978, mikheyev1985}, spectral swaps \cite{duan2006, duan2006b, raffelt2007, raffelt2007c, fogli2007, dasgupta2009, friedland2010, galais2011}, and matter--neutrino resonances (MNRs) \cite{malkus2012, malkus2014, malkus2016, wu2016, vaananen2016, zhu2016}. Quantum adiabaticity is the key concept in all three cases. Quantum-adiabatic MSW resonances, spectral swaps, and MNRs are adiabatic in the thermodynamic sense as well: they are processes in which the medium does work on the system without heating it. As examples, we will show explicitly how miscidynamics reproduces adiabatic MSW conversion and spectral swaps.

Consider neutrinos propagating radially outward from a static stellar environment with equation of motion
\begin{equation}
    i \partial_r \rho_\nu = [H_\nu, \rho_\nu],
\end{equation}
where $\rho_\nu$ and $H_\nu$ are functions of radial coordinate $r$ and neutrino energy $E_\nu$. The Hamiltonian is independent of $\rho_\nu$ because neutrino--neutrino forward scattering is taken to be negligible, as in the standard MSW effect. In this collisionless scenario, the adiabatic miscidynamic equation is simply
\begin{equation}
    i \partial_r \rho_\nu^\textrm{eq} = 0, \label{eq:miscMSW}
\end{equation}
with $\rho_\nu^{\textrm{eq}}$ satisfying $[ H_\nu, \rho_\nu^{\textrm{eq}}] = 0$. An ``eq'' superscript on $H_\nu$ is unnecessary because of the Hamiltonian's lack of dependence on $\rho_\nu$. The local-equilibrium condition is equivalent to
\begin{equation}
    \vec{H}_\nu \times \vec{P}_\nu^{\textrm{eq}} = 0,
\end{equation}
which implies that $\vec{P}_\nu^{\textrm{eq}}$ is aligned with $\vec{H}_\nu$ at all $r$. Both $P_{\nu,0}$ and $|\vec{P}_\nu|$ are constant because $iC_\nu = 0$. Therefore Eq.~\eqref{eq:miscMSW} implies the adiabatic MSW effect. Due to the coarse-graining, $\vec{P}^{\textrm{eq}}_\nu$ must be aligned with $H_\nu$, whereas in the fine-grained MSW effect the relative angle between the two vectors needs to be fixed. Alignment emerges from the fine-grained dynamics because the coarse-graining averages over many precession periods of $\vec{P}_\nu$ about $\vec{H}_\nu$.

Spectral swaps were analyzed in Refs.~\cite{raffelt2007, raffelt2007c} using the equation of motion
\begin{equation}
    i \partial_t \rho_\nu = [H_\nu, \rho_\nu],
\end{equation}
with the evolution as a function of $t$ instead of $r$. The Hamiltonian in this case includes the self-coupling part $H_\nu^{(2)}$. Since the system is again collisionless, the adiabatic miscidynamic equation is virtually the same as before,
\begin{equation}
    i \partial_t \rho_\nu^\textrm{eq} = 0, \label{eq:miscswap}
\end{equation}
with the important distinction entering through the local-equilibrium condition $[H^{\textrm{eq}}_\nu, \rho^{\textrm{eq}}_\nu] = 0$, which now must be solved self-consistently because of the dependence of $H_\nu$ on $\rho_\nu$. As before, instantaneous mixing equilibrium implies that alignment is maintained between $\vec{P}_\nu^{\textrm{eq}}$ and $\vec{H}_\nu^{\textrm{eq}}$, and the absence of collisions entails the constancy of $P_{\nu,0}$ and $|\vec{P}_\nu|$. Under the assumption that every $\vec{P}_\nu^{\textrm{eq}}$ is initially aligned with its respective $\vec{H}_\nu^{\textrm{eq}}$, adiabatic miscidynamics requires
\begin{equation}
    \vec{P}_\nu^{\textrm{eq}} = |\vec{P}_\nu^{\textrm{eq}}| \hat{H}_\nu^{\textrm{eq}}. \label{eq:alignswap}
\end{equation}
Yet this is precisely the condition that is used in Refs.~\cite{raffelt2007,raffelt2007c} to obtain analytic spectral-swap solutions. The self-consistency conditions that Ref.~\cite{raffelt2007} derives from Eq.~\eqref{eq:alignswap} [Eqs.~(11) and (13) of that work] are equivalent to the self-consistency conditions in Eq.~\eqref{eq:selfcons} above. Therefore miscidynamics reproduces adiabatic spectral swaps.

\subsection{Equilibration}

Other well-known oscillation phenomena are instances of heat-generating mixing equilibration. The most basic example is the kinematic decoherence of neutrinos (\textit{i.e.}, the averaging-out of their oscillations) after traveling many oscillation lengths. As shown below, maximizing the entropy of a neutrino in vacuum produces \begin{equation}
    \rho^\textrm{eq}_{ij} = \rho^\textrm{in}_{ii} \delta_{ij}, \label{eq:vacanswer}
\end{equation}
where $\rho^\textrm{in}$ is the distribution at the source. Neutrinos decohere in the mass basis, which ordinarily we would attribute to wave-packet separation \cite{akhmedov2009}. Here we recognize it as thermalization. The calculation is simple but illustrative.

We use $\overline{\rho}$ to indicate the coarse-grained density matrix $\rho$. The particular coarse-graining we employ is inessential. It could be an average over a window of energies, for example, which would correspond to an experiment in which we make finite-resolution detections of neutrinos in a beam of average energy $p$. For this subsection, we will simply assume that there is \textit{some} coarse-graining---some conduit by which information about the system is lost---that justifies the maximization of entropy.

Because the neutrinos are in vacuum, the Hamiltonian is
\begin{equation}
H = p + M^2 / 2p.
\end{equation}
The conserved quantities in this scenario are $\textrm{Tr}\left[ \overline{\rho} \right]$ and $\textrm{Tr}\left[ H \overline{\rho} \right]$. The nonlinear quantity $\textrm{Tr}[\overline{\rho}^2]$ is not conserved because $\overline{\rho^2} \neq \overline{\rho}^2$. The statistical ensemble therefore has expectation values
\begin{align}
&\textrm{Tr} \left[ \overline{\rho} \right] = \overline{\textrm{Tr}\left[ \rho \right]} \equiv \mathcal{N}, \notag \\
&\textrm{Tr}\left[ H \overline{\rho} \right] = \overline{\textrm{Tr}\left[ H \rho \right]} \equiv \mathcal{U}. \label{eq:vacconslaws}
\end{align}

Let $\eta$ and $\beta$ be the Lagrange multipliers associated with $\mathcal{N}$ and $\mathcal{U}$, respectively. Then the entropy with constraints is
\begin{align}
S' = &- \textrm{Tr} \left[ \overline{\rho} \log \overline{\rho} + \left( 1 - \overline{\rho} \right) \log \left( 1 - \overline{\rho} \right) \right] \notag \\
&- \eta \left( \textrm{Tr} \left[ \overline{\rho} \right] - \mathcal{N} \right) + \beta \left( \textrm{Tr} \left[ H \overline{\rho} \right] - \mathcal{U} \right).
\end{align}
From
\begin{equation}
\frac{\delta S'}{\delta \overline{\rho}} \bigg|_{\rho^\textrm{eq}} = 0
\end{equation}
we obtain
\begin{equation}
\rho^\textrm{eq} = \frac{1}{\exp\left[ \beta \left( p + \frac{M^2}{2p} \right)  - \eta\right] + 1}. \label{eq:vaceq}
\end{equation}
The distribution is diagonal in the mass basis. Working in this basis, the conservation laws in Eq.~\eqref{eq:vacconslaws} give
\begin{equation}
\sum_{i=1}^{N_f} \left( p + \frac{m_i^2}{2p} \right)^n \rho_{ii}^\textrm{eq} = \sum_{i=1}^{N_f} \left( p + \frac{m_i^2}{2p} \right)^n \rho_{ii}^\textrm{in}
\end{equation}
for $n = 0, 1$. Hence the problem is solved by Eq.~\eqref{eq:vacanswer} as claimed.

This example is simple enough that we can find the solution without knowing $\eta$ and  $\beta$, but an explicit determination of the Lagrange multipliers could be carried out numerically if desired. To do that, we would solve the self-consistency conditions
\begin{align}
&\mathcal{N} = \textrm{Tr} \left[ \rho^\textrm{eq} \left( \eta, \beta \right) \right], \notag \\
&\mathcal{U} = \textrm{Tr} \left[ H \rho^\textrm{eq} \left( \eta, \beta \right) \right],
\end{align}
which result from mandating agreement between the conservation laws in Eq.~\eqref{eq:vacconslaws} and the functional form of $\rho^\textrm{eq}$ in Eq.~\eqref{eq:vaceq}.

Incidentally, a similar calculation shows up in the thermodynamics of massless neutrinos that are isolated from their surroundings but interact with each other through $2 \rightarrow 2$ collisions. For a single neutrino species, the conserved quantities would be temperature $T$ and chemical potential $\mu$, and they would have to be obtained by solving the self-consistency relations among number density $n$, energy density $u$, and the Fermi--Dirac distribution $f_\textrm{FD} (p, T, \mu)$:
\begin{align}
&n = \int \frac{d^3 \boldsymbol{p}}{(2\pi)^3} f_\textrm{FD} \left( p, T, \mu \right), \notag \\
&u = \int \frac{d^3 \boldsymbol{p}}{(2\pi)^3} p f_\textrm{FD} \left( p, T, \mu \right).
\end{align} 
These types of calculations, where one must infer intensive thermodynamic parameters from known extensive ones, arise in considering the thermalization of isolated systems because such systems act as their own environments.

Collective flavor instabilities are the means by which \textit{self-interacting} neutrino systems transition from an unstable to a stable mixing equilibrium. To illustrate the power of this viewpoint, let us consider fast instabilities, which grow on $\mathcal{O}(\textrm{cm})$ length scales and are pervasive in simulations of mergers and supernovae \cite{tamborra2021, richers2022}. Over several years and scores of studies, we have come to understand that fast instabilities occur if and only if neutrino distributions exhibit a certain type of angular crossing \cite{sawyer2016, chakraborty2016c, dasgupta2017, izaguirre2017, capozzi2017, abbar2018, capozzi2019, martin2020, johns2021, morinaga2022, dasgupta2022}---but we have continued to lack a physical explanation why. Thermodynamics offers one: angular crossings make it possible for $S$ to increase while fixing $\int d^3\boldsymbol{x}~D_0 (t, \boldsymbol{x})$, which is invariant in the usual models of fast flavor conversion. Instabilities epitomize the ergodic maxim that anything that can happen, will. This perspective also comports with the insufficiency of angular crossings to ensure instability in homogeneous systems, where additional conservation laws inhibit the dynamics \cite{johns2020}. In point of fact, homogeneous flavor evolution resembles the dynamics of a finite mechanical system \cite{hannestad2006, duan2007b, raffelt2013b, johns2018, johns2020, padillagay2022, fiorillo2023, fiorillo2023b} and is not expected to be thermodynamic in character.

Numerical experiments have repeatedly shown that instabilities lead to phase-space cascades and quasi-steady states \cite{sawyer2005, raffelt2007b, mangano2014, mirizzi2015, johns2020b, bhattacharyya2020, richers2021b, nagakura2022}. Recently the focus has been on characterizing the asymptotic outcomes of fast instabilities in particular \cite{bhattacharyya2021, nagakura2023b, zaizen2023b, zaizen2023, xiong2023d, richers2024asymptotic, george2024evolution, abbar2024physics, liu2025asymptotic}. Thermodynamics posits that the observed states fluctuate around mixing equilibria, and predicts the mean distributions to take the form of Eq.~\eqref{eq:rhoeq} with suitably chosen invariants. This maximum-entropy hypothesis can be tested by checking whether post-instability asymptotic states do in fact show polarization vectors $\vec{P}_\nu$ with the dependence on $\boldsymbol{p}$ that is predicted by Eq.~\eqref{eq:rhoeq}. We will not undertake numerical tests here, but we do note that $\rho_\nu^\textrm{eq}$ is implied by ergodicity with little other input. If numerical observations turn out not to agree with the proposed distribution, that in itself will be interesting.

Collisions bring about equilibration as well, but towards a distinct equilibrium. If we were to turn on collisions in one of our periodic boxes, they would gradually cause flavor depolarization and heating. But we have chosen $l_\textrm{box}$ and $t_\textrm{box}$ to be small enough that these effects are minor. As a result, collisions operate entirely at the coarse-grained level. Their effects take place during the transit \textit{between} boxes. In this manner we separate mixing equilibration, which by assumption is complete, and neutrino--fluid equilibration, which is not.

\section{Conclusion}

We have applied thermodynamics to coarse-grained neutrino density matrices. This theory affords a new way of understanding the phenomenology of neutrino flavor conversion. Motivated by statistical principles, we have also proposed the adiabatic limit of miscidynamics, a transport theory based on the assumption that neutrinos maintain local mixing equilibrium.

In deriving adiabatic miscidynamics, we assumed subgrid fluctuations to be uncorrelated. The nonadiabatic extension of miscidynamics, which can accommodate the influence of subgrid correlations on the grid-level evolution, will be presented elsewhere \cite{johns2025local}.

\begin{acknowledgments}
This work greatly benefited from the insight and encouragement of Carlos Arg\"{u}elles, Adam Burrows, Vincenzo Cirigliano, Huaiyu Duan, Damiano Fiorillo, Bei-Lok Hu, Anson Kost, Hiroki Nagakura, Georg Raffelt, Sanjay Reddy, Alessandro Roggero, Meng-Ru Wu, and Zewei Xiong, and from the hospitality of the Mainz Institute for Theoretical Physics (MITP) of the Cluster of Excellence PRISMA+ (Project ID 39083149). Support for this work was provided by NASA through Hubble Fellowship grant number HST-HF2-51461.001-A awarded by the Space Telescope Science Institute, which is operated by the Association of Universities for Research in Astronomy, Incorporated, under NASA contract NAS5-26555, and by a Feynman Fellowship through LANL LDRD project number 20230788PRD1.
\end{acknowledgments}

\bibliography{all_papers}

\end{document}